\documentclass[conference]{IEEEtran}
\IEEEoverridecommandlockouts

\usepackage{cite}
\usepackage{algorithm}
\usepackage[noend]{algpseudocode}
\usepackage{amsmath,amssymb,amsfonts}
\usepackage{graphicx}
\usepackage{subcaption}
\usepackage{textcomp}
\usepackage{xcolor}
\usepackage{pifont}
\usepackage{multirow}
\usepackage{makecell}
\usepackage{array}

\begin{document}

\title{Semantic Communications for Speech Signals}
\author{Zhenzi Weng\IEEEauthorrefmark{1}, Zhijin Qin\IEEEauthorrefmark{1}, Geoffrey Ye Li\IEEEauthorrefmark{2}\\
\small \IEEEauthorrefmark{1} Queen Mary University of London, London, UK\\
\small \IEEEauthorrefmark{2} Imperial College London, London, UK\\
\small \{zhenzi.weng, z.qin\}@qmul.ac.uk, geoffrey.li@imperial.ac.uk
}

\maketitle

\begin{abstract}
We consider a semantic communication system for speech signals, named DeepSC-S. Motivated by the breakthroughs in deep learning (DL), we make an effort to recover the transmitted speech signals in the semantic communication systems, which minimizes the error at the semantic level rather than the bit level or symbol level as in the traditional communication systems. Particularly, based on an attention mechanism employing squeeze-and-excitation (SE) networks, we design the transceiver as an end-to-end (E2E) system, which learns and extracts the essential speech information. Furthermore, in order to facilitate the proposed DeepSC-S to work well on dynamic practical communication scenarios, we find a model yielding good performance when coping with various channel environments without retraining process. The simulation results demonstrate that our proposed DeepSC-S is more robust to channel variations and outperforms the traditional communication systems, especially in the low signal-to-noise (SNR) regime.
\end{abstract}


\IEEEpeerreviewmaketitle

\section{Introduction}
Intelligent communications have been investigated recently to break though the bottlenecks of the traditional communication systems\cite{b1}. Inspired by the success of deep learning (DL) in various areas, such as computer vision and natural language processing (NLP), DL has been considered as a promising candidate to revolutionize communication systems with high performance and intelligence\cite{b2}. Particularly, DL has shown its great potentials to solve the existing technical problems in physical layer communications\cite{b3,b4,b5} and wireless resource allocations\cite{b6,b7}.

Even if the communication systems utilizing DL technique yield better performance than the traditional approaches for some scenarios and conditions, most of the literature focus on the performance improvement at the bit or symbol level, which usually takes bit-error rate (BER) or symbol-error rate (SER) as the performance metrics. Particularly, the major task in the traditional communication systems and the developed DL-enabled systems, is to recover the transmitted message accurately and effectively, represented by digital bit sequences. In the past decades, such type of wireless communication systems have experienced significant development from the first generation (1G) to the fifth generation (5G) and the system capacity is approaching Shannon limit. Based on Shannon and Weaver\cite{b8}, communications can be categorized into three levels as follow: \romannumeral1)  \emph{level A}: how accurately can the symbols of communication be transmitted? \romannumeral2) \emph{level B}: how precisely do the transmitted symbols convey the desired meaning? \romannumeral3) \emph{level C}: how effectively does the received meaning affect conduct in the desired way? This indicates the feasibility to transmit the semantic information, instead of the bits or symbols, to achieve higher system efficiency.

Semantic information, in contrast to information theory exploited in existing communication systems, takes into account the meaning and veracity of source information because it can be both informative and factual\cite{b9}, which facilitates the semantic communication systems to recover information via further utilizing the meaning difference between the input and the recovered signals\cite{b10}. According to the recent efforts in\cite{b11}, semantic data can be compressed to proper size for transmission using a lossless method by utilizing the semantic relationship between different messages, while the traditional lossless source coding is to represent a signal with the minimum number of binary bits by exploring the dependencies or statistical properties of input signals. In addition, inspired by the E2E communication systems\cite{b12}, different types of sources have been considered in recent investigations on E2E semantic communication systems, which mainly focus on the image and text transmission\cite{b13,b14,b15,b16,b17,b18,b19,b20}. The investigation on semantic communications for speech signals transmission is still missed.

Particularly, an initial research on semantic communication systems for text information has been developed\cite{b13}, which mitigates the semantic error to achieve Nash equilibrium. However, such a text-based semantic communication system only measures the performance at the word level instead of the sentence level. Thus, a further investigation about semantic communications for text transmission, named DeepSC, has been carried out\cite{b14} to deal with the semantic error at the sentence level with various length. Moreover, a lite distributed semantic communication system for text transmission, named L-DeepSC, has been proposed\cite{b15} to address the challenge of IoT to perform the intelligent tasks.

In the area of semantic communications for image information, a DL-enabled semantic communication system for image transmission, named JSCC, has been developed\cite{b16}. Based on JSCC, an image transmission system, integrating channel output feedback, has been investigated to improve image reconstruction\cite{b17}. Similar to text transmission, IoT applications for image transmission have been carried out. Particularly, a joint image transmission-recognition system has been developed\cite{b18} to achieve high recognition accuracy and a deep joint source-channel coding architecture, name DeepJSCC, has been investigated\cite{b19} to process image with low computation complexity. 

In this article, we explore the semantic systems for speech signals by utilizing DL technique. Particularly, a DL-enabled semantic communication system for speech signals, named DeepSC-S, is proposed to address the existing challenges in the traditional communication systems, e.g., the block-wise system has been demonstrated to be sub-optimal, and conventional linear signal processing algorithm is unable to capture many imperfections and non-linearities in the practical channel. The main contributions of this article can be summarized as threefold:

\begin{itemize}
\item A novel semantic communication system for speech signals, named DeepSC-S, is first proposed, which treats the whole transceiver as two deep neural networks (DNNs), and jointly designs the speech coding and the channel coding to deal with source distortion and channel effects.

\item Particularly, in the proposed DeepSC-S, the squeeze-and-excitation (SE) networks\cite{b20} is employed to learn and extract the essential speech semantic information, as well assign high values to the weights corresponding to the essential information during the training phase. By exploiting the attention mechanism based on SE networks, DeepSC-S improves the accuracy of signal recovering.

\item Moreover, by training DeepSC-S under a fixed fading channel and SNR, then facilitating the trained model with good performance under testing channel conditions, the proposed DeepSC-S is highly robust to dynamic channel environments without network tuning and retraining.
\end{itemize}

The rest of this article is structured as follows. Section \uppercase\expandafter{\romannumeral2} introduces the model of speech semantic communication system and performance metrics. In Section \uppercase\expandafter{\romannumeral3}, the details of the proposed DeepSC-S is presented. Simulation results are discussed in Section \uppercase\expandafter{\romannumeral4} and Section \uppercase\expandafter{\romannumeral5} draws conclusions.

\section{System Model}
In this section, we first introduce the considered system model. Besides, the details of the system model and the performance metrics are presented. 

\subsection{System Expectation}
\begin{figure}[tbp]
\centering 
\includegraphics[width=0.45\textwidth]{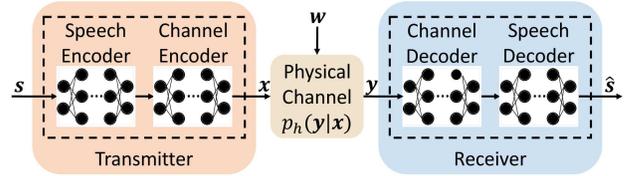} 
\caption{The model structure of DL-enabled semantic communication system for speech signals.}
\label{sys model}
\end{figure}

The considered system will transmit the original speech signals via a neural network (NN)-based speech semantic communication system, which comprises two major tasks as shown in Fig. \ref{sys model}: \romannumeral1) semantic information learning and extracting of speech signals; \romannumeral2) and mitigating the effects of wireless channels. Due to the variation of speech characteristics, it is a quite challenging problem. For a practical communication scenario, the signal passing through the physical channel suffers from distortion and attenuation. Therefore, the considered DL-enabled system targets to recover the original speech signals and achieve better performance than the traditional approaches while coping with complicated channel distortions.

\subsection{Transmitter}
The proposed system model is shown in Fig. \ref{sys model}. From the figure, the input of the transmitter is a speech sample sequence, $\boldsymbol s=\left[s_1,\;s_2,\;...,\;s_W\right]$ with $W$ samples, where $s_w$ is $w$-th item in $\boldsymbol s$ and it is a scalar value, i.e., a positive number, a negative number, or zero. At the transmitter, the input, $\boldsymbol s$, is mapped into symbols, $\boldsymbol x$, to be transmitted over physical channels. As shown in Fig. \ref{sys model}, the transmitter consists of two individual components: the \emph{speech encoder} and the \emph{channel encoder}, in which each component is implemented by an independent NN. Denote the NN parameters of the \emph{speech encoder} and the \emph{channel encoder} as $\boldsymbol\alpha$ and $\boldsymbol\beta$, respectively. Then the encoded symbol sequence, $\boldsymbol x$, can be expressed as
\begin{equation}
\boldsymbol x=\mathbf T_{\boldsymbol\beta}^{\mathcal C}(\mathbf T_{\boldsymbol\alpha}^{\mathcal S}(\boldsymbol s)),
\label{auto-encoder}
\end{equation}
where $\mathbf T_{\boldsymbol\alpha}^{\mathcal S}(\cdot)$ and $\mathbf T_{\boldsymbol\beta}^{\mathcal C}(\cdot)$ indicate the \emph{speech encoder} and the \emph{channel encoder} with respect to (w.r.t.) parameters $\boldsymbol\alpha$ and $\boldsymbol\beta$, respectively. Here we denote the NN parameters of the transmitter as $\boldsymbol\theta^{\mathcal T}=(\boldsymbol\alpha,\boldsymbol\;\boldsymbol\beta)$.

The mapped symbols, $\boldsymbol x$, are transmitted over a physical channel. Note that the normalization on transmitted symbols $\boldsymbol x$ is required to ensure the total transmission power constraint $\mathbb{E}\left\|\boldsymbol x\right\|^2=1$.

The whole transceiver in Fig. \ref{sys model} is designed for a single communication link, in which the channel layer, represented by $p_h\left(\left.\boldsymbol y\right|\boldsymbol x\right)$, takes $\boldsymbol x$ as the input and produces the output as received signal, $\boldsymbol y$. Denote the coefficients of a linear channel as $\boldsymbol h$, then the transmission process from the transmitter to the receiver can be modeled as
\begin{equation}
\boldsymbol y=\boldsymbol h\ast\boldsymbol x+\boldsymbol w,
\label{channel}
\end{equation}
where $\boldsymbol w\sim\mathcal{CN}(0,\;\sigma^2\mathbf I)$ indicates independent and identically distributed (i.i.d.) Gaussian noise, $\sigma^2$ is noise variance for each channel and $\mathbf I$ is the identity matrix. 

\subsection{Receiver}
Similar to the transmitter, the receiver also consists of two cascaded parts, including the \emph{channel decoder} and the \emph{speech decoder}. The \emph{channel decoder} is to mitigate the channel distortion and attenuation, and the \emph{speech decoder} recovers speech signals based on the learned and extracted speech semantic features. Denote the NN parameters of the \emph{channel decoder} and the \emph{speech decoder} as $\boldsymbol\chi$ and $\boldsymbol\delta$, respectively. As depicted in Fig. \ref{sys model}, the decoded signal, $\widehat{\boldsymbol s}$, can be obtained from the received signal, $\boldsymbol y$, by the following operation:
\begin{equation}
\widehat{\boldsymbol s}=\mathbf R_{\boldsymbol\delta}^{\mathcal S}(\mathbf R_{\boldsymbol\chi}^{\mathcal C}(\boldsymbol y)),
\label{auto-decoder}
\end{equation}
where $\mathbf R_{\boldsymbol\chi}^{\mathcal C}(\cdot)$ and $\mathbf R_{\boldsymbol\delta}^{\mathcal S}(\cdot)$ indicate the \emph{channel decoder} and the \emph{speech decoder} w.r.t. parameters $\boldsymbol\chi$ and $\boldsymbol\delta$, respectively. Denote the NN parameter of the receiver as $\boldsymbol\theta^{\mathcal R}=(\boldsymbol\chi,\boldsymbol\;\boldsymbol\delta)$.

The objective of the whole transceiver system is to recover the speech information as close as to the original. In the traditional communication systems, the performance is achieved at the bit level to target a low BER/SER. However, in our proposed system, the bit-to-symbol transformation is not involved. Thus, for the sake of forming a powerful loss function, we treat the speech recovery process after going though the whole transceiver as a signal reconstruction task by exploiting the characteristics of speech signals, then mean-squared error (MSE) can be used as the loss function in our system to measure the difference between $\boldsymbol s$ and $\widehat{\boldsymbol s}$, denoted as
\begin{equation}
{\mathcal L}_{MSE}(\boldsymbol\theta^{\mathcal T},\;\boldsymbol\theta^{\mathcal R})=\frac1W\sum_{w=1}^W{(s_w-{\widehat s}_w)}^2.
\label{loss function}
\end{equation}

\subsection{Performance Metrics}
In our model, the system is committed to reconstruct the raw speech signals. Hence, the signal-to-distortion ration (SDR)\cite{b21} is employed to measure the ${\mathcal L}_2$ error between $\boldsymbol s$ and $\widehat{\boldsymbol s}$, which can be expressed as
\begin{equation}
SDR=10\log_{10}\left(\frac{\left\|\boldsymbol s\right\|^2}{\left\|\boldsymbol s-\widehat{\boldsymbol s}\right\|^2}\right).
\label{SDR}
\end{equation}
The higher SDR represents the speech information is recovered with better quality, i.e., easier to understand for human beings. According to (\ref{loss function}), MSE loss could reflect the goodness of SDR. The lower the MSE, the higher the SDR.

Perceptual evaluation of speech distortion (PESQ)\cite{b22} is considered as another metric to measure the quality of listening at the receiver, which assumes the short memory in human perception. PESQ is a speech quality assessment model combing the perceptual speech quality measure (PSQM) and perceptual analysis measurement system (PAMS), which is in International Telecommunication Union (ITU-T) recommendation P.862. PESQ is a good candidate for evaluating the quality of speech messages under various conditions, e.g., background noise, analog filtering, and variable delay, by scoring the speech quality range from -0.5 to 4.5.

\section{Proposed Semantic Communication System for Speech Signals}
To address the aforementioned challenges, we design a DL-enabled speech semantic communication system, named DeepSC-S. Specifically, an attention-based two-dimension (2D) CNN is used for the speech coding and a 2D CNN is adopted for the channel coding. The details of the developed DeepSC-S will be introduced in this section.

\subsection{Model Description}
\begin{figure*}[htbp]
\includegraphics[width=1\textwidth]{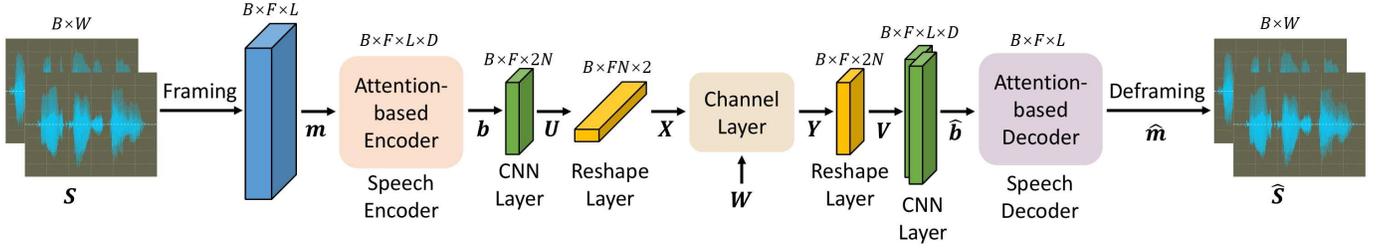} 
\centering 
\caption{The proposed system architecture for semantic communication system for speech signals.}  
\label{proposed sys}  
\end{figure*}

As shown in Fig. \ref{proposed sys}, the input of the proposed DeepSC-S, denoted as $\boldsymbol S\in\mathfrak R^{B\times W}$, is the set of speech sample sequences, $\boldsymbol s$, which are drawn from the speech dataset, $\mathfrak S$, and $B$ is the batch size. The input sample sequences set, $\boldsymbol S$, are framed into $\boldsymbol m\in\mathfrak R^{B\times F\times L}$ for training before passing through an attention-based encoder, i.e., the \emph{speech encoder}, where $F$ indicates the number of frames and $L$ is the length of each frame. The \emph{speech encoder} directly learns the speech semantic information from $\boldsymbol m$ and outputs the learned features $\boldsymbol b\in\mathfrak R^{B\times F\times L\times D}$. Afterwards, the \emph{channel encoder}, denoted as a CNN layer with 2D CNN modules, converts $\boldsymbol b$ into $\boldsymbol U\in\mathfrak R^{B\times F\times2N}$. In order to transmit $\boldsymbol U$ into a physical channel, it is reshaped into symbol sequences, $\boldsymbol X\in\mathfrak R^{B\times FN\times2}$, via a reshape layer.

The channel layer takes the reshaped symbol sequences, $\boldsymbol X$, as the input and produces $\boldsymbol Y$ at the receiver, which is given by 
\begin{equation}
\boldsymbol Y=\boldsymbol H\boldsymbol X+\boldsymbol W,
\label{proposed channel}
\end{equation}
where $\boldsymbol H$ consists of $B$ number of channel coefficient vectors, $\boldsymbol h$, and $\boldsymbol W$ is Gaussian noise, which includes $B$ number of noise vectors, $\boldsymbol w$.

The received symbol sequences, $\boldsymbol Y$, is reshaped into $\boldsymbol V\in\mathfrak R^{B\times F\times2N}$ before feeding into the \emph{channel decoder}, represented by a CNN layer with 2D CNN modules. The output of the \emph{channel decoder} is $\widehat{\boldsymbol b}\in\mathfrak R^{B\times F\times L\times D}$. Afterwards, an attention-based decoder, i.e., the \emph{speech decoder}, converts $\widehat{\boldsymbol b}$ into $\widehat{\boldsymbol m}\in\mathfrak R^{B\times F\times L}$ and $\widehat{\boldsymbol m}$ is recovered into $\widehat{\boldsymbol S}$ via the inverse operation of framing, named deframing, where the size of $\widehat{\boldsymbol S}$ is same as that of $\boldsymbol S$ at the transmitter. The loss is calculated at the end of the receiver and backpropagated to the transmitter, thus, the trainable parameters in the whole system can be updated simultaneously.

\subsection{Speech Encoder and Decoder}
\begin{figure}[htbp]
\includegraphics[width=0.48\textwidth]{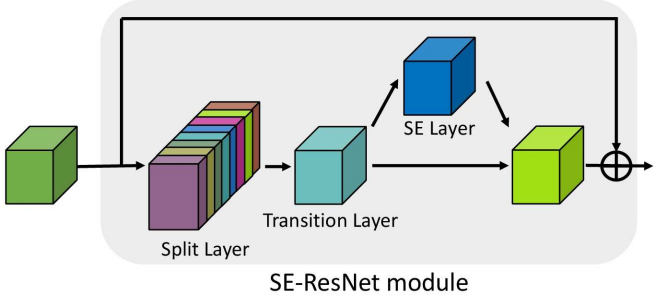} 
\centering 
\caption{SE-ResNet module for constructing the \emph{speech encoder} and the \emph{speech decoder}.}  
\label{SE-ResNet}  
\end{figure}

The core of the proposed DeepSC-S is the NN-enabled \emph{speech encoder} and \emph{speech decoder} based on an attention mechanism, named SE-ResNet, as shown in Fig. \ref{SE-ResNet}, where the SE layer is considered as an independent unit and one or multiple SE-ResNet modules are sequentially connected to constrct the \emph{speech encoder} and the \emph{speech decoder}.

The motivation of attention mechanism is to identify the essential information and the weights corresponding to the essential information are assigned to high values when weight updating and adjusting during the training phase. In this work, we focus on learning the speech semantic information, such as the stressed speech signals. Particularly, for SE networks, a \emph{squeeze} operation is first implemented to aggregate the 2D spatial dimension of each input feature, then an operation, named \emph{excitation}, intents to learn and output the attention factor of each feature by capturing the inter-dependencies. Accordingly, the weights of input to SE-ResNet are reassigned, i.e., the weights corresponding to the essential speech information are paid more attention. Additionally, residual network is adopted to alleviate the problem of gradient vanishing due to the network depth.

\subsection{Model Training and Testing}
Based on the prior knowledge of channel state information (CSI), the transmitter and receiver parameters, $\boldsymbol\theta^{\mathcal T}$ and $\boldsymbol\theta^{\mathcal R}$, can be updated simultaneously. As aforementioned, the objective of the proposed DeepSC-S is to train a model to capture the essential information in speech signals and make it to work well under various channels and a wide SNR regime.

\subsubsection{Training Stage} 
As in Fig. \ref{proposed sys}, the training algorithm of DeepSC-S is described in Algorithm \ref{training algorithm}. During the training stage, in order to achieve a valid training task, the MSE loss converges until the loss is no longer decreasing. The number of SE-ResNet modules is an important hyperparameter, which aims to facilitate the good performance of the \emph{speech encoder/decoder}, as well as the reasonable training time.
\begin{algorithm}[htb]
\caption{Training algorithm of the proposed DeepSC-S.}
\label{training algorithm}
\textbf{Initialization:} initialize parameters $\boldsymbol\theta^{\mathcal T{\boldsymbol(\mathbf0\boldsymbol)}}$ and $\boldsymbol\theta^{\mathcal R{\boldsymbol(\mathbf0\boldsymbol)}}$, $i=0$.

\begin{algorithmic}[1]
    \State \textbf{Input:} Speech sample sequences $\boldsymbol S$ from speech dataset $\mathfrak S$, a fading channel $\boldsymbol h$, noise $\boldsymbol w$ generated under a fixed SNR value.
    \State Framing $\boldsymbol S$ into $\boldsymbol m$ with trainable size.
        \While{Stop criterion is not meet}
            \State $\mathbf T_{\boldsymbol\beta}^{\mathcal C}(\mathbf T_{\boldsymbol\alpha}^{\mathcal S}(\boldsymbol m))\rightarrow\boldsymbol X$.
            \State Transmit $\boldsymbol X$ over physical channel and receive $\boldsymbol Y$ via \State (\ref{channel}).
            \State $\mathbf R_{\boldsymbol\delta}^{\mathcal S}(\mathbf R_{\boldsymbol\chi}^{\mathcal C}(\boldsymbol Y))\rightarrow\widehat{\boldsymbol m}$.
            \State Deframing $\widehat{\boldsymbol m}$ into $\widehat{\boldsymbol S}$.
            \State Compute loss ${\mathcal L}_{MSE}(\boldsymbol\theta^{\mathcal T},\;\boldsymbol\theta^{\mathcal R})$ via (\ref{loss function}).
            \State Update trainable parameters simultaneously via SGD:
                \begin{equation}
                \boldsymbol\theta^{\mathcal T\left(i+1\right)}\leftarrow\boldsymbol\theta^{\mathcal T\left(i\right)}-\eta\nabla_{\boldsymbol\theta^{\mathcal T\left(i\right)}}{\mathcal L}_{MSE}(\boldsymbol\theta^{\mathcal T},\;\boldsymbol\theta^{\mathcal R})
                \label{transmitter paramter update}
                \end{equation}
                \begin{equation}
                \boldsymbol\theta^{\mathcal R\left(i+1\right)}\leftarrow\boldsymbol\theta^{\mathcal R\left(i\right)}-\eta\nabla_{\boldsymbol\theta^{\mathcal R\left(i\right)}}{\mathcal L}_{MSE}(\boldsymbol\theta^{\mathcal T},\;\boldsymbol\theta^{\mathcal R})
                \label{receiver paramter update}
                \end{equation}
            \State $i\leftarrow i+1$.
        \EndWhile
    \State \textbf{end while}
    \State \textbf{Output:} Trained networks $\mathbf T_{\boldsymbol\alpha}^{\mathcal S}(\cdot)$, $\mathbf T_{\boldsymbol\beta}^{\mathcal C}(\cdot)$, $\mathbf R_{\boldsymbol\chi}^{\mathcal C}(\cdot)$, and $\mathbf R_{\boldsymbol\delta}^{\mathcal S}(\cdot)$.
\end{algorithmic}

\end{algorithm}

\subsubsection{Testing Stage}
Based on the trained networks $\mathbf T_{\boldsymbol\alpha}^{\mathcal S}(\cdot)$, $\mathbf T_{\boldsymbol\beta}^{\mathcal C}(\cdot)$, $\mathbf R_{\boldsymbol\chi}^{\mathcal C}(\cdot)$, and $\mathbf R_{\boldsymbol\delta}^{\mathcal S}(\cdot)$ from the outputs of Algorithm \ref{training algorithm}, the testing algorithm of DeepSC-S is illustrated in Algorithm \ref{testing algorithm}. As shown in Algorithm \ref{testing algorithm}, the trained model under a fixed channel condition is employed to test the performance under various fading channels directly without model retraining.
\begin{algorithm}[htb]
\caption{Testing algorithm of the proposed DeepSC-S.}
\label{testing algorithm}

\begin{algorithmic}[1]   
    \State \textbf{Input:} Speech sample sequences $\boldsymbol S$ from speech dataset $\mathfrak S$, trained networks $\mathbf T_{\boldsymbol\alpha}^{\mathcal S}(\cdot)$, $\mathbf T_{\boldsymbol\beta}^{\mathcal C}(\cdot)$, $\mathbf R_{\boldsymbol\chi}^{\mathcal C}(\cdot)$, and $\mathbf R_{\boldsymbol\delta}^{\mathcal S}(\cdot)$, testing channel set $\mathcal H$, a wide range of SNR regime.
    \State Framing $\boldsymbol S$ into $\boldsymbol m$ with trainable size.
    	\For{each channel condition $\boldsymbol h$ drawn from $\mathcal H$}
    	    \For{each SNR value}
    	        \State Generate Gaussian noise $\boldsymbol w$ under the SNR value.
    	        \State $\mathbf T_{\boldsymbol\beta}^{\mathcal C}(\mathbf T_{\boldsymbol\alpha}^{\mathcal S}(\boldsymbol m))\rightarrow\boldsymbol X$.
                \State Transmit $\boldsymbol X$ over physical channel and receive $\boldsymbol Y$ \State via (\ref{channel}).
                \State $\mathbf R_{\boldsymbol\delta}^{\mathcal S}(\mathbf R_{\boldsymbol\chi}^{\mathcal C}(\boldsymbol Y))\rightarrow\widehat{\boldsymbol m}$.
                \State Deframing $\widehat{\boldsymbol m}$ into $\widehat{\boldsymbol S}$.
                \EndFor
            \State \textbf{end for}
        \EndFor
    \State \textbf{end for}
	\State \textbf{Output:} Recovered speech sample sequences, $\widehat{\boldsymbol S}$, under different fading channels and various SNR values.
\end{algorithmic}

\end{algorithm}

\section{Experiment and Numerical Results}
In this section, we compare to the performance between the proposed DeepSC-S and the traditional communication systems for speech signals transmission over telephone systems under the AWGN channels, the Rayleigh channels, and the Rician channels, where the accurate CSI is assumed.
\begin{figure*}[tbp]
\begin{minipage}[t]{0.33\linewidth}
\centering
\includegraphics[width=1\textwidth]{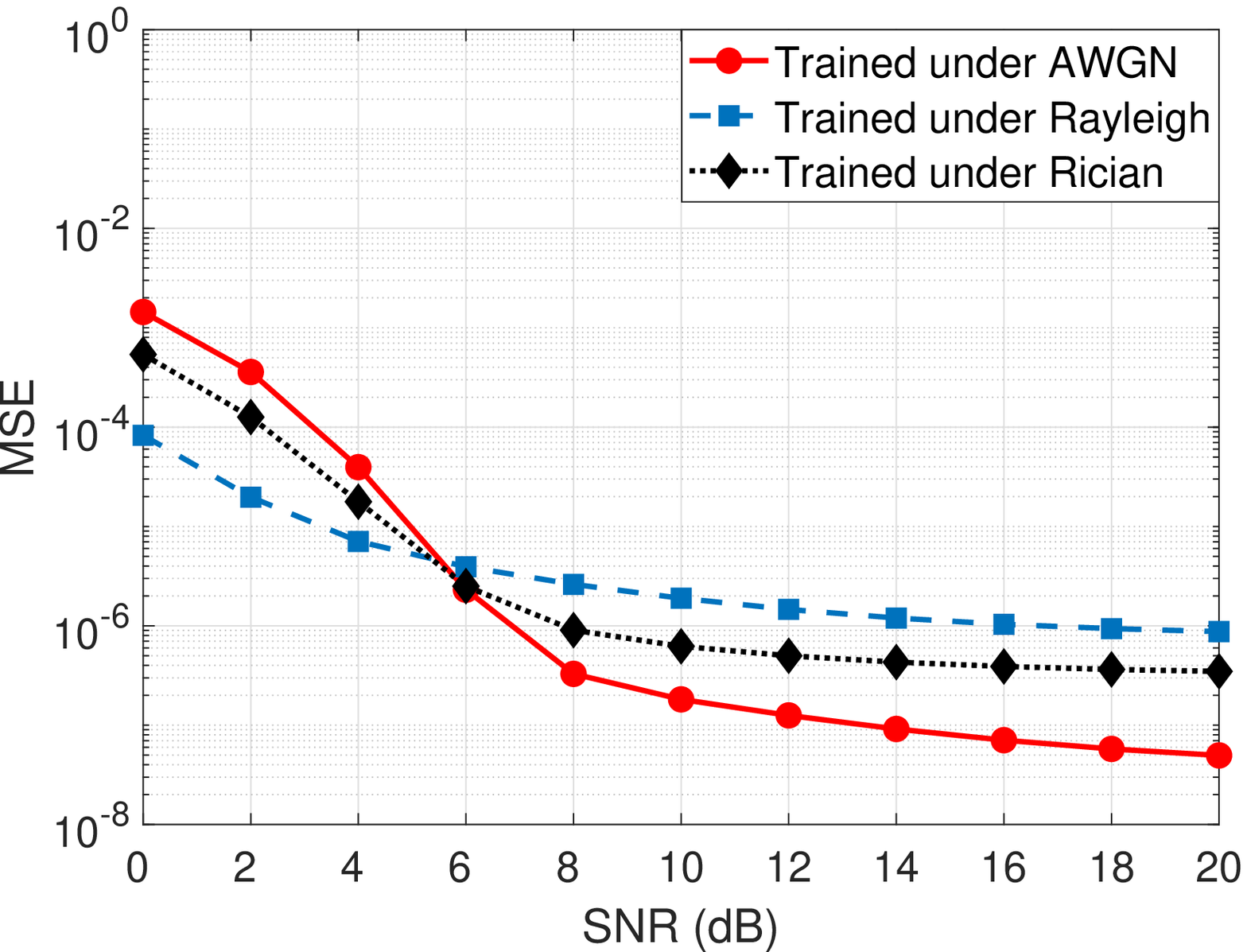} 
\subcaption{AWGN channels}
\label{MSE tested under AWGN channel}
\end{minipage}
\begin{minipage}[t]{0.33\linewidth}
\centering
\includegraphics[width=1\textwidth]{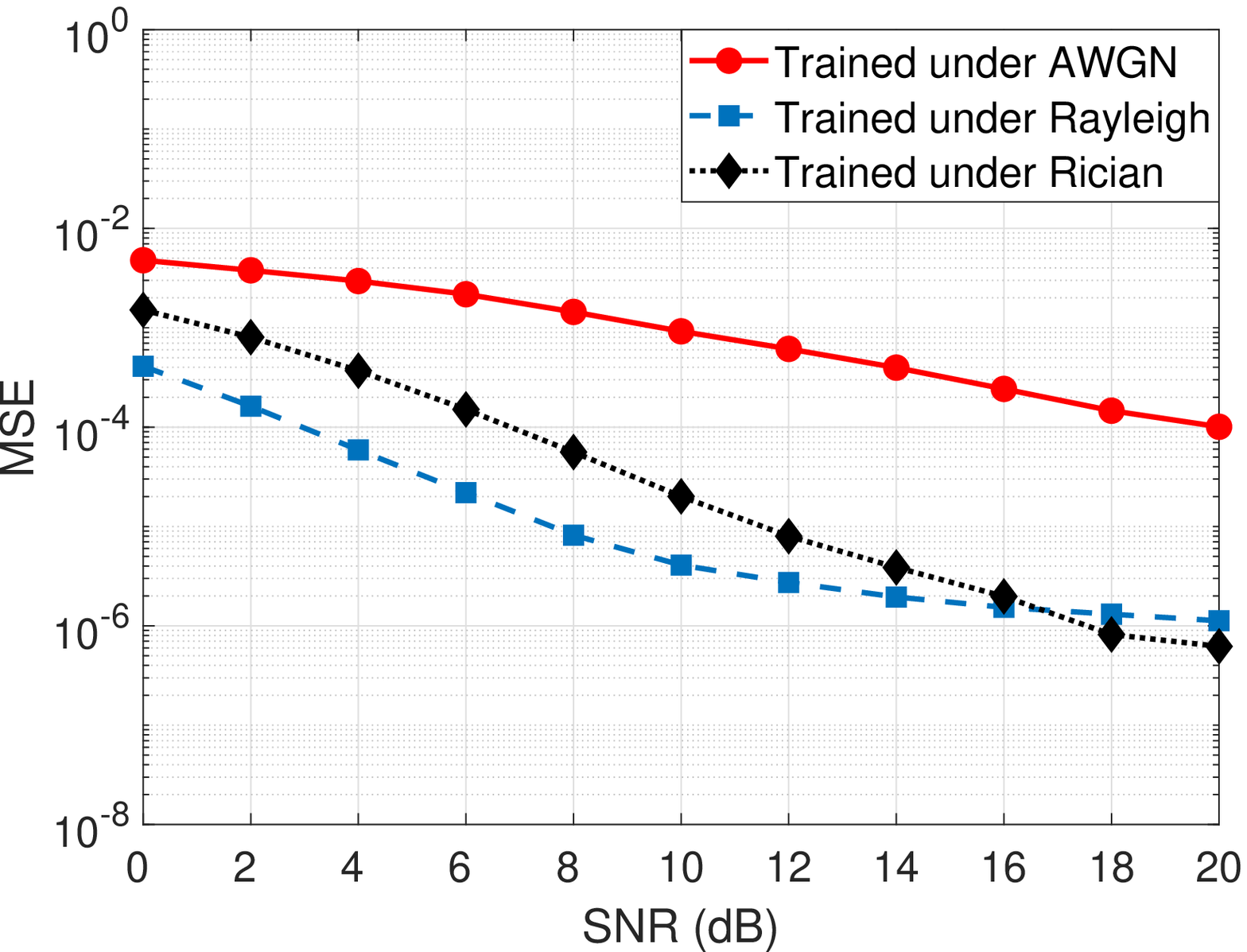} 
\subcaption{Rayleigh channels}
\label{MSE tested under Rayleigh channel}
\end{minipage} 
\begin{minipage}[t]{0.33\linewidth}
\centering
\includegraphics[width=1\textwidth]{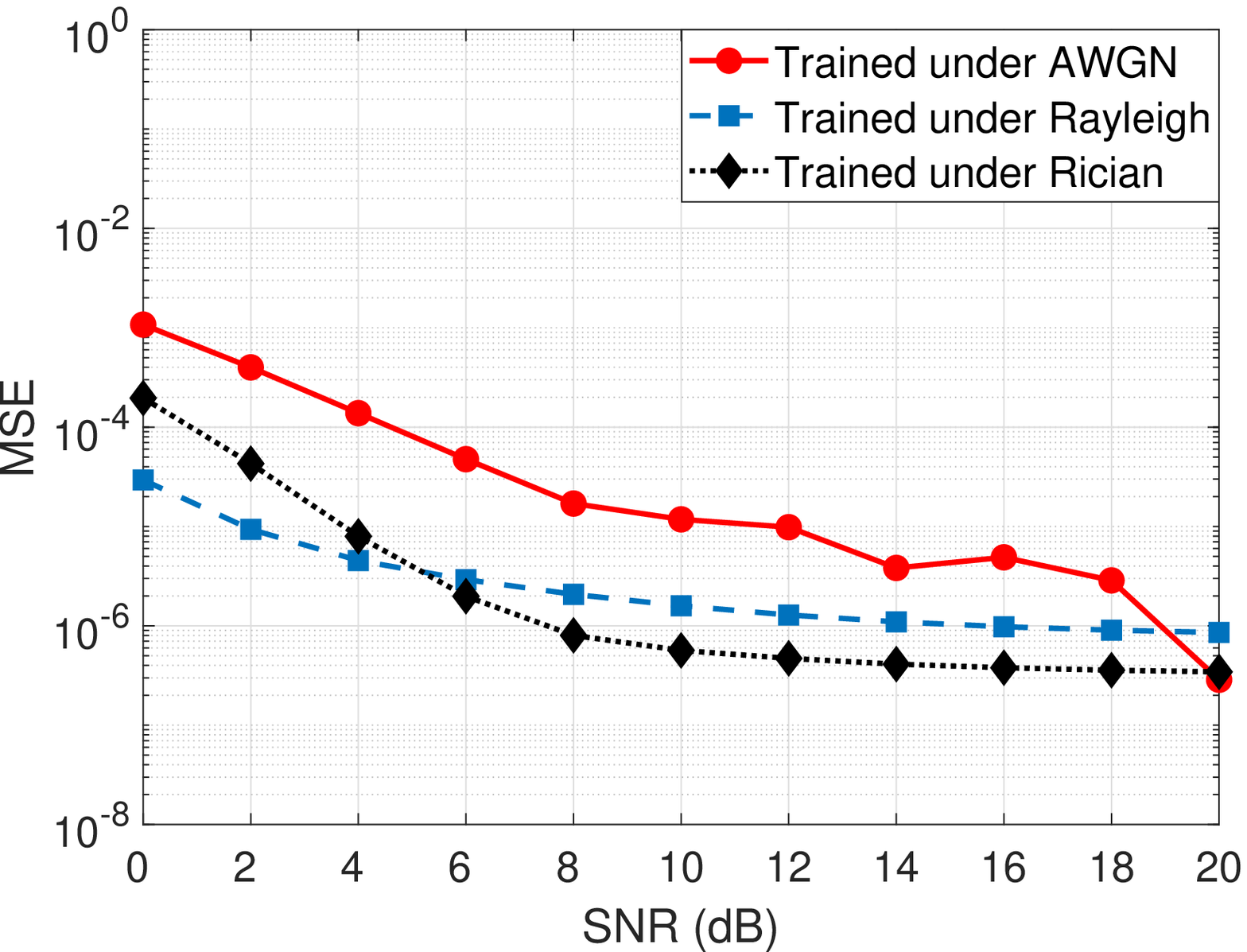}
\subcaption{Rician channels}
\label{MSE tested under Rician channel}
\end{minipage} 
\caption{MSE loss tested for (a) AWGN, (b) Rayleigh, and (c) Rician channels with the models trained under various channels.}
\label{MSE comparison for different testing channels}
\end{figure*}

\subsection{Speech Dataset and Traditional Model}
In the whole experiment, we adopt the speech dataset from Edinburgh DataShare, which comprises more than 10,000 \emph{.wav} files trainset and 800 \emph{.wav} files testset with sampling rate 16KHz. In terms of the traditional telephone systems, the sampling rate for speech signals is 8KHz, thus, the speech samples are down-sampled to 8KHz. Note that the number of speech samples in different \emph{.wav} files is inconsistent. In the simulation, we fix $W=16,384,$ and each sample sequence in $\boldsymbol m$ consists of frames $F=128$ with the frame length $L=128$.

According to ITU-T G.711 standard, 64 Kbps pulse code modulation (PCM) is recommended for speech source coding in telephone systems with $2^8=256$ quantization levels\cite{b23}. For the channel coding, turbo codes with soft output Viterbi algorithm (SOVA) is considered to improve the performance of error detection and correction at the receiver\cite{b24}, in which the coding rate is 1/3, the block length is 512, and the number of decoding iterations is 5. In addition, to make the number of transmitted symbols in the traditional systems is same as that in DeepSC-S, 64-QAM is adopted in the traditional systems for the modulation.

\subsection{Experiments over Telephone Systems}
\subsubsection{A Robust Model}
In this experiment, we investigate a robust system to work on various channel conditions by training DeepSC-S under the fixed channel condition, and then testing the MSE loss via the trained model under all adopted fading channels. Particularly, the number of the SE-ResNet modules in the \emph{speech encoder/decoder} is 4 and the number of the 2D CNN modules in the \emph{channel encoder/decoder} is 1, which includes 8 kernels. The network setting of the proposed DeepSC-S are shown as Table \ref{telephone DeepSC-S NN parameters}.
\renewcommand\arraystretch{1.15} 
\begin{table}[htbp]
\footnotesize
\caption{Parameters settings of the proposed DeepSC-S for telephone systems.}
\label{telephone DeepSC-S NN parameters}
\centering
\begin{tabular}{|c|c|c|c|}
\hline
               & \textbf{Layer Name} & \textbf{Kernels} & \textbf{Activation}\\
\hline
    \multirow{2}{6.0em}{\textbf{\centering \ Transmitter}} & 4$\times$SE-ResNet & 4$\times$32   & Relu \\
\cline{2-4}
                                                                         & CNN layer   &    8   &   Relu    \\
\cline{1-4}
    \multirow{3}{6.0em}{\textbf{\centering \ \ \ Receiver}}   & CNN layer   &    8   &   Relu    \\
\cline{2-4}
                                    & 4$\times$SE-ResNet      &    4$\times$32    &   Relu     \\
\cline{2-4}
                                    & 1$\times$CNN module     &   1               &   None     \\
\cline{1-4}
    \textbf{Learning Rate}          & $\eta$                  &    0.001          &   None     \\
\hline
\end{tabular}
\end{table}

As shown in Fig. \ref{MSE comparison for different testing channels} (a), in terms of the MSE loss tested under the AWGN channels, DeepSC-S trained under the AWGN channels outperforms the model trained under the Rayleigh channels and the Rician channels when SNRs are higher than around 6 dB. Besides, according to Fig. \ref{MSE comparison for different testing channels} (b), DeepSC-S trained under the AWGN channels performs quite poor in terms of MSE loss when testing under the Rayleigh channels. Furthermore, Fig. \ref{MSE comparison for different testing channels} (c) shows the model trained under the three adopted channels can achieve MSE loss values under $9\times10^{-7}$ when testing under the Rician channels. Therefore, DeepSC-S trained under the Rician channels is considered as a robust model that is capable of coping with various channel environments.

Note that during the training stage, the Gaussian noise in three channels are generated under a fixed SNR value, 8 dB. According to Fig. \ref{MSE comparison for different testing channels}, when SNR in three testing channels is lower than 8 dB, DeepSC-S trained under the AWGN channels has higher MSE loss values than the model trained under Rayleigh channels and the Rician channels.

\subsubsection{SDR and PESQ Results}
Based on the robust model, i.e., DeepSC-S trained under the Rician channels and 8 dB SNR, we test the SDR and PESQ under DeepSC-S and the traditional systems for speech transmission over telephones systems.

\begin{figure}[htbp]
\includegraphics[width=0.45\textwidth]{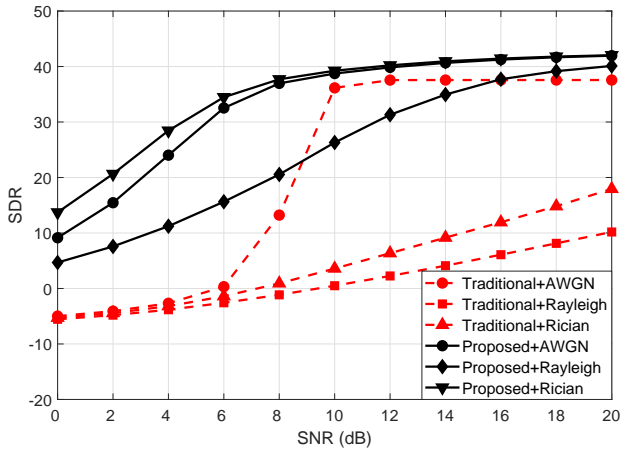} 
\centering 
\caption{SDR score versus SNR for the traditional speech communication systems with 8-bits A-law PCM coding with Turbo codes in 64-QAM and DeepSC-S under the AWGN channels, the Rayleigh channels, and the Rician channels.}  
\label{SDR result for speech communcation}  
\end{figure}

Fig. \ref{SDR result for speech communcation} compares the SDR performance between DeepSC-S and the traditional communication systems under the AWGN channels, the Rayleigh channels, and the Rician channels, which shows that DeepSC-S achieves better SDR than the traditional one under all tested channels. Moreover, DeepSC-S performs steadily when coping with different channels and SNRs, while for the traditional model, its performance is quite poor under dynamic channel conditions, especially in the low SNR regime, DeepSC-S significantly outperforms the traditional systems. Furthermore, DeepSC-S yields higher SDR scores under the Rician channels than the AWGN channels because the model is trained under the Rician channels.

\begin{figure}[htbp]
\includegraphics[width=0.45\textwidth]{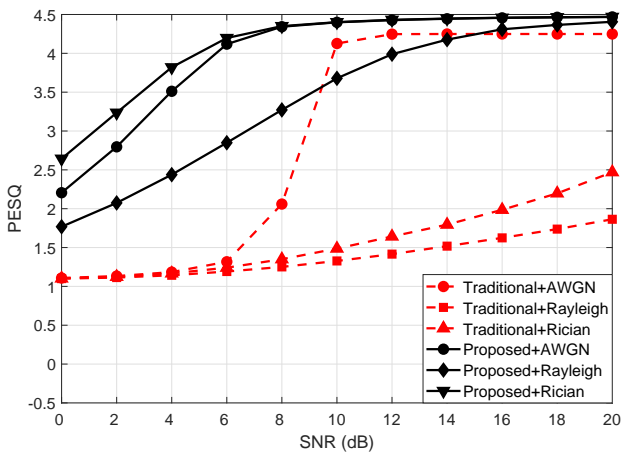} 
\centering 
\caption{PESQ score versus SNR for the traditional speech communication systems with 8-bits A-law PCM coding with Turbo codes in 64-QAM and DeepSC-S  under the AWGN channels, the Rayleigh channels, and the Rician channels.}  
\label{PESQ result for speech communcation}  
\end{figure}

The PESQ score comparison is in Fig. \ref{PESQ result for speech communcation}. From the figure, the proposed DeepSC-S can provide high quality speech recovery and outperforms the traditional approaches under various fading channels and SNRs. Moreover, similar to the results of SDR, DeepSC-S obtains good PESQ scores when coping with channel variations while the traditional one provides poor scores under the low SNR regime. According to the simulation results, DeepSC-S is able to yield better speech transmission service in the complicated communication scenarios than the traditional systems, especially in the low SNR regime.

\section{Conclusions}
In this article, we investigate a DL-enabled semantic communication system for speech signals, named DeepSC-S, which achieves more efficient transmission than the traditional systems by utilizing the speech semantic information. Particularly, we jointly design the speech coding and the channel coding to learn and extract the essential speech information. Additionally, an attention mechanism based on squeeze-and-excitation (SE) networks is utilized to improve the recovery accuracy. Moreover, in order to facilitate DeepSC-S working well over various physical channels, a model with strong robustness to channel variations is investigated. Simulation results demonstrated that DeepSC-S outperforms the traditional communication systems, especially in the low SNR regime. Hence, our proposed DeepSC-S is a promising candidate for speech semantic communication systems.

\end{document}